\documentclass[aps,prb,twocolumn,showpacs,superscriptaddress]{revtex4-2} 

\usepackage{amsmath,amssymb,mathrsfs}
\usepackage{latexsym}
\usepackage{graphicx} 
\usepackage{epstopdf}
\usepackage{graphicx,epstopdf,color}
\usepackage{amsfonts}
\usepackage{bm}
\usepackage{multirow}
\usepackage{dsfont}
\usepackage[
    colorlinks=true,
    linkcolor=blue,
    citecolor=blue,
    urlcolor=blue
]{hyperref}
\usepackage{CJK}

\begin{document}
\begin{CJK*}{UTF8}{gbsn}	

\title{Fractional vortices in a spin-isotropic spiral spin liquid}

\author{Cecilie Glittum}
\affiliation{Department of Physics, University of Oslo, P. O. Box 1048 Blindern, N-0316 Oslo, Norway}
\affiliation{Helmholtz-Zentrum Berlin f\"ur Materialien und Energie, Hahn-Meitner Platz 1, 14109 Berlin, Germany}
\affiliation{Dahlem Center for Complex Quantum Systems and Fachbereich Physik, Freie Universit\"at Berlin, Arnimallee 14, 14195 Berlin, Germany}

\author{Han Yan (闫寒)}
\affiliation{Institute for Solid State Physics, The University of Tokyo, Kashiwa, Chiba 277-8581, Japan}

\author{Johannes Reuther}
\affiliation{Helmholtz-Zentrum Berlin f\"ur Materialien und Energie, Hahn-Meitner Platz 1, 14109 Berlin, Germany}
\affiliation{Dahlem Center for Complex Quantum Systems and Fachbereich Physik, Freie Universit\"at Berlin, Arnimallee 14, 14195 Berlin, Germany}

\date{\today}

\begin{abstract}
Spiral spin liquids are magnetic states whose classical ground-state manifold consists of planar incommensurate spin spirals with wave vectors lying on a continuous ring or surface in reciprocal space. The resulting subextensive degeneracy suppresses magnetic long-range order and gives rise to liquid-like behavior. Despite numerous material realizations and theoretical investigations, the structure of low-temperature spin configurations of spin-isotropic Heisenberg spiral spin liquids has remained poorly understood. Here, we classify and characterize the classical topological defects supported by these systems. We uncover a rich family of vortex types involving concerted windings of spin directions, spiral-plane normals, and wave-vector orientations, yielding a $\mathds{Z}\times\mathds{Z}_2$ classification. Remarkably, the elementary defects are half-vortices carrying fractional $2\pi$ windings in both spin and momentum space and obey fusion rules resembling to those of Ising anyons.  Large-scale classical simulations of a square-lattice spiral spin liquid reveal that these vortices are dense in the spiral-spin-liquid regime, and bind tightly below an order-by-disorder phase transition, eventually fusing to vacuum.
\end{abstract}
\maketitle
\end{CJK*}

\section{Introduction}\label{sec:intro}

Spin liquids are strongly correlated magnetic phases that evade conventional magnetic long-range order down to temperatures well below the dominant exchange scale~\cite{Anderson1973,Balents2010,SavaryBalents2017,Broholm2020}. Their low-energy physics is instead controlled by collective constraints and, in quantum spin liquids, by long-range entanglement and emergent gauge structure, giving rise to fractionalized quasiparticles and topological phenomena. Consequently, their characterization rests not on a local order parameter alone, but on the structure of correlations, topological defects, and emergent gauge fields. Spin liquids therefore stand at a central stage for investigating collective phenomena beyond the conventional symmetry-breaking paradigm.

Spiral spin liquids are frustrated magnets whose low-energy states form a continuous manifold of spin spirals. Rather than selecting one of finitely many propagation vectors, competing interactions generate a contour in two-dimensional reciprocal space or a surface in three dimensions~\cite{Bergman2007,Mulder2010,Okumura2010,Yao2021}. Correlations remain distributed along this manifold, producing characteristic rings or surfaces in the static structure factor and soft collective fluctuations while preserving global magnetic disorder. This coexistence of local spiral structure and global propagation-vector degeneracy distinguishes a spiral spin liquid from both a conventional paramagnet and an ordered spiral magnet.

Spiral-spin-liquid behavior has now been reported across wide classes of materials. Among three-dimensional magnets, $\mathrm{MnSc}_2\mathrm{S}_4$ realizes a spiral surface~\cite{Krimmel2006,Gao2017MnSc2S4,Gao2020MnSc2S4}, whereas $\mathrm{LiYbO}_2$ and $\mathrm{Cs}_3\mathrm{Fe}_2\mathrm{Cl}_9$ host one-dimensional spiral contours embedded in three-dimensional reciprocal space~\cite{Bordelon2021,Graham2023,Gao2026Cs3Fe2Cl9}. Quasi-two-dimensional Heisenberg systems include $\mathrm{FeCl}_3$, $\mathrm{AgCrSe}_2$ and $\mathrm{Ca}_{10}\mathrm{Cr}_7\mathrm{O}_{28}$~\cite{Gao2022,Baenitz2021,Andriushin2025,Balz2016,Pohle2021,Takahashi2025}, while $\mathrm{GdZnPO}$ and $\mathrm{CaMn}_2\mathrm{P}_2$ provide easy-plane counterparts~\cite{Wan2024,Zhao2025Magnetocaloric,Zhao2025Transport,Chen2026,Islam2025CaMn2P2}. A spiral ring has also been reported in the metallic system $\mathrm{EuAg}_4\mathrm{Sb}_2$~\cite{Neves2026EuAg4Sb2}, extending spiral-manifold physics beyond insulating local-moment magnets.

Theory has established broad lattice and interaction criteria for spiral manifolds to appear and characterized their correlations and thermodynamics~\cite{Attig2017,Buessen2018,Niggemann2020,Yao2021,Glittum2021,Gao2022LineGraph,Liu2022,Yan2024Classification,Glittum2026}. Such states have also been connected to helicoidal and multiple-$q$ textures, vector chirality, meron and skyrmion configurations, and smectic-like elasticity~\cite{Nattermann_2018,Seabra2016,Shimokawa2019Ripple,Shimokawa2019MultipleQ,Glittum2021,Huang2022,Mohylna2022,Mohylna2025,Hsieh2023}. Much less is known about the intrinsic topological defects of the liquid itself. In the easy-plane, or XY, limit, winding of the local propagation vector produces unconventional momentum vortices constrained by the compatibility of the spin angle~\cite{Yan2022,Gonzalez2024}. Most experimental candidates, however, are at or close to the Heisenberg limit, for which the spiral plane can rotate in spin space and the defect structure must incorporate both the propagation-vector orientation and the spin frame.

Here we develop the theory of topological defects and associated phases of a spin-isotropic Heisenberg spiral spin liquid. The local order-parameter manifold is $[\mathrm{U}(1)\times \mathrm{SO}(3)]/\mathbb{Z}_2$, whose quotient couples reversal of the propagation vector to reversal of the spiral-plane normal. This permits an elementary defect combining a half winding of the spiral ring with a spin-frame rotation, yielding two chiralities; conventional momentum and spin-frame vortices are composites. The resulting topological defects are classified by $\mathbb{Z}\times\mathbb{Z}_2$ quantized numbers. Large-scale Monte Carlo (MC) simulations identify these defects in the correlated liquid regime, establishing a topological framework for dynamics and textures in Heisenberg spiral spin liquids. Our results have direct implications for the known spiral spin liquid materials, making them promising candidates for hosting these fractional topological defects.

\section{Results}\label{sec:results}

\subsection{The model}\label{sec:model}
\begin{figure*}[t]
    \centering
    \includegraphics[width = \linewidth]{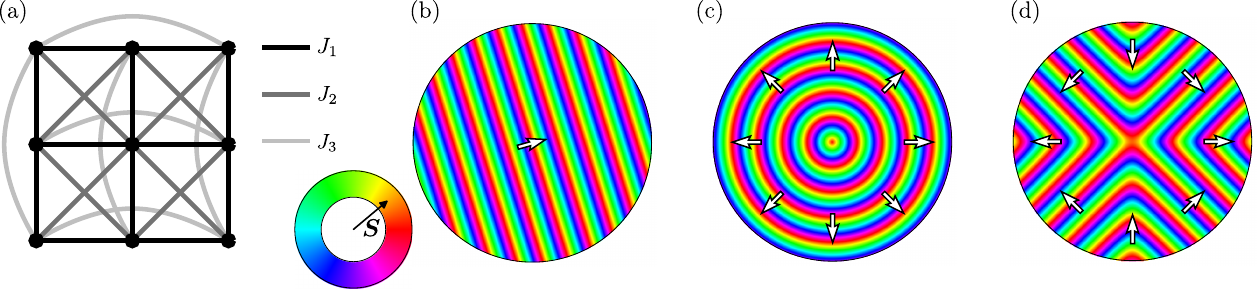}
    \caption{(a) Definition of couplings. (b) Homogeneous planar spiral. (c) Planar momentum vortex with $u'=1$. (d) planar momentum vortex with $u'=-1$. Hollow arrows indicate the local momentum vectors.}\label{fig:XY-spiral}
\end{figure*}

We consider the following classical Heisenberg Hamiltonian on the square lattice
\begin{equation}
\mathcal{H}=J_1\sum_{\langle i,j\rangle_1}{\bm S}_i\cdot{\bm S}_j+J_2\sum_{\langle i,j\rangle_2}{\bm S}_i\cdot{\bm S}_j+J_3\sum_{\langle i,j\rangle_3}{\bm S}_i\cdot{\bm S}_j,\label{eq:ham}
\end{equation}
where $\langle i,j\rangle_n$ are pairs of $n$th neighbor sites (counted only once) coupled by the exchange interactions $J_n$ up to third neighbors $n=3$, see Fig.~\ref{fig:XY-spiral}(a).  ${\bm S}_i$ are normalized ($|{\bm S}_i|=1$) classical three-component spin vectors. Throughout this paper we consider the case
\begin{equation}
J_1=-1,\quad J_2>1/4,\quad J_3=J_2/2.
\end{equation}
These interactions are chosen such that the classical ground states are planar spin spirals of the form
\begin{equation}
{\bm S}_i={\bm e}_1 \cos({\bm q}\cdot{\bm r}_i+\phi)+{\bm e}_2 \sin({\bm q}\cdot{\bm r}_i+\phi)\label{eq:plain_spiral}
\end{equation}
where ${\bm r}_i$ is the position of site $i$, the angle $\phi$ is a global spiral phase and ${\bm e}_1$ and ${\bm e}_2$ are two orthogonal unit vectors that span the spiral plane, which can be chosen arbitrarily for a spin-isotropic Heisenberg model. One can show that, to be a ground state, the spiral wave vector $\bm q$ has to fulfill the condition
\begin{equation}
\cos (q_x)+\cos(q_y)=\frac{1}{2J_2},\label{eq:spiral_ring}
\end{equation}
forcing ${\bm q}$ to lie on a ring-like manifold in reciprocal space, whose size grows with increasing $J_2$. In the limit of small $\delta\equiv J_2-1/4\ll1$ these rings acquire a circular shape with radius $q_\circ\approx4\sqrt{\delta}$. In the following we are primarily interested in the physical properties in this limit, where the circular ring shape push conventional thermal order-by-disorder effects that would select specific types of magnetic order on the ring down to low temperatures. We thus set $\delta=0.03$ in the following but emphasize that our results are independent of model details and could also be found for spiral-spin-liquid models on honeycomb or triangular lattices, as long as the spiral ring is sufficiently circular to prevent order-by-disorder effects at temperature of interest~\cite{Villain1980, Henley1989, Chandra1990}.

\subsection{Properties of the XY model}\label{sec:xy}
We first briefly review the properties of the corresponding XY model with two-component spins ${\bm S}_i=(S_i^x,S_i^y)$, as previously investigated in Ref.~\cite{Yan2022}, and then generalize it to the Heisenberg model with three-component spins. Spin directions in the $x$-$y$ plane are parametrized by a single polar angle $\Phi$ such that the spiral ground state in Eq.~(\ref{eq:plain_spiral}) can be written as
\begin{equation}
{\bm S}_i=(\cos(\Phi({\bm r}_i)),\sin(\Phi({\bm r}_i))),\label{eq:xy_spiral}
\end{equation}
with $\Phi({\bm r}_i)={\bm q}\cdot{\bm r}_i+\phi$ where ${\bm q}$ again has to fulfill the condition in Eq.~(\ref{eq:spiral_ring}), see Fig.~\ref{fig:XY-spiral}(b) for a homogeneous planar ground state spiral. Considering the continuum limit ${\bm r}_i\rightarrow {\bm r}$ it directly follows ${\bm q}={\bm \nabla}\Phi(\bm r)$ where ${\bm \nabla}$ is the two-dimensional gradient.

Small deformations of exact spiral ground states are described by a wave-vector field ${\bm q}={\bm q}({\bm r})$ whose direction can fluctuate spatially and whose magnitude can slightly deviate from the ground-state value, i.e., $|{\bm q}({\bm r})|\approx q_\circ$. Importantly, due to its property of being a gradient field, the wave vector is curl-free, ${\bm \nabla}\times{\bm q}(\bm r)=0$.

As discussed in Ref.~\cite{Yan2022} the system's order parameter space is formed by two subspaces, a U(1) degree of freedom that corresponds to a global spin-rotation symmetry in the $x$-$y$-plane and another wave-vector related U(1) subspace that corresponds to the freedom of ${\bm q}$ to occupy any point on the spiral ring. 
%While the wave vector degree of freedom is not associated with a global symmetry across all energy scales it can nevertheless be considered as an effective {\it low-energy symmetry}. 
The standard homotopy group argument then implies that each U(1) order parameter space gives rise to topological defects characterized by a $\mathds{Z}$-valued winding number, but the physics is more complicated than that. 
First, the system can host conventional spin vortices which correspond to singular points in the wave vector field ${\bm q}(\bm r)$. 
In numerical simulations they emerge as small but energetically costly defects. Second, the wave-vector U(1) degree of freedom likewise gives rise to a vortex, dubbed {\it momentum vortex} in Ref.~\cite{Yan2022}. This unconventional topological defect does not require non-analytic points in ${\bm q}(\bm r)$, however, the additional curl-free condition ${\bm \nabla}\times{\bm q}(\bm r)=0$ puts constraints on the allowed momentum winding numbers $u'$. Specifically, it was shown that $u'$ can only take values $u'=1,0,-1,-2,\ldots$. We show examples for momentum vortices with $u'=1$ and $u'=-1$ in Fig.~\ref{fig:XY-spiral}(c) and (d). At low temperature the system's dynamics was found to be primarily governed by fluctuations through momentum vortex configurations which tend to become increasingly rigid and network-like.
\begin{figure*}[t]
    \centering
    \includegraphics[width = 0.99\linewidth]{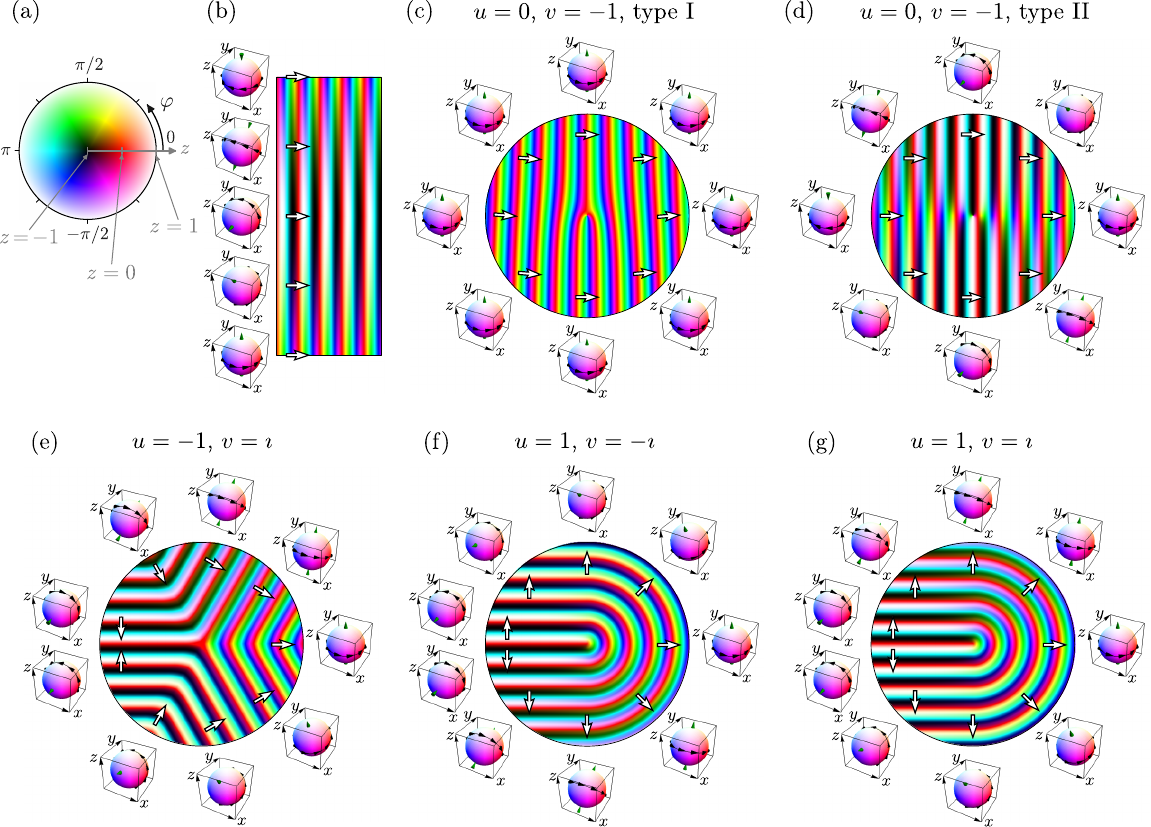}
    \caption{(a) Definition of the color scale for the spin directions in the following subfigures: spins have rainbow colors along the equator ($x$-$y$-plane, parametrized by the polar angle $\varphi$) getting lighter (darker) when moving towards the north pole at $z=1$ (south pole at $z=-1$). (b) Non-coplanar deformation of a spiral state where the spins rotate out of the $x$-$y$ plane when moving vertically in real space. This non-coplanar angle variation is described by the smaller eigenvalue $\lambda_-$ of $G_{\mu\nu}$. Here and in the following subfigures, the spheres indicate the variations of the spin directions along the arrows on the black ring, when moving in the direction of the nearby hollow arrows (momentum vectors) in real space. Green arrows correspond to the normal direction of the spiral ${\bm \Omega}_q$. (c) $\mathds{Z}_2$ vortex of type I with $u=0$, $v=-1$ where $\bm S$ performs a net $2\pi$ rotation while ${\bm \Omega}_q$ remains fixed. (d) $\mathds{Z}_2$ vortex of type-II with $u=0$, $v=-1$ where ${\bm \Omega}_q$ performs a $2\pi$ rotation. (e) Half-momentum antivortex with $u=1$ and $v=\imath$. (f) and (g) both illustrate half-momentum vortices with $u=1$ which differ by their chirality $v=-\imath$ and $v=\imath$, respectively.}\label{fig:vortices}
\end{figure*}

For the discussion of the Heisenberg system it will be helpful to use a generalized  definition of the wave vector field ${\bm q}(\bm r)$ that is based on the $2\times2$ tensor
\begin{equation}
G_{\mu\nu}(\bm r)=\left[\partial_\mu {\bm S}(\bm r)\right]\cdot \left[\partial_\nu {\bm S}(\bm r)\right],\label{eq:g_tensor}
\end{equation}
where $\mu,\nu\in\{x,y\}$. Inserting Eq.~(\ref{eq:xy_spiral}) yields $G_{\mu\nu}=q_\mu q_\nu$ which has eigenvalues $\lambda_+={\bm q}^2$ and $\lambda_-=0$ with the corresponding normalized eigenvectors ${\bm v}_+={\bm q}/|{\bm q}|$ and ${\bm v}_-=(q_y,-q_x)/|{\bm q}|$. 
Consequently, the wave vector ${\bm q}$ follows from the larger eigenvalue and the corresponding eigenvector through ${\bm q}=\sqrt{\lambda_+}{\bm v}_+$, while the other eigenvalue always vanishes in the planar XY case. 
Physically, the eigenvectors of $G_{\mu\nu}$ identify the real-space directions along which the spin angle varies most rapidly and most slowly. The direction of fastest variation is set by the wave vector, whereas the slowest direction is perpendicular to it (parallel to the spiral wavefronts) along which the spin angle remains constant.

\subsection{Vortex types of the Heisenberg system}\label{sec:vortices}
With these preparations we now investigate the Heisenberg case where we first specify the order parameter space which then allows us to characterize the topological defects.
\subsubsection{Order-parameter space}
Like in the XY-case the order parameter space has a spin and a wave-vector part. For an investigation of the latter, we first need to find a proper definition of the wave vector suitable for three-component spins where small deformations of exact spiral ground states may not be coplanar. In this case, the definition of the wave vector ${\bm q}={\bm \nabla}\Phi(\bm r)$ can no longer be applied since a parametrization with a single spin angle is not possible. Here, we generalize the definition of the wave vector via Eq.~(\ref{eq:g_tensor}). While in the XY-case $\partial_x {\bm S}$ and $\partial_y {\bm S}$ are parallel, both oriented along the direction $(-\sin(\Phi),\cos(\Phi))$ in spin space, in the Heisenberg case $\partial_x {\bm S}$ and $\partial_y {\bm S}$ need not be parallel, implying that the spirals are not locally planar. In terms of an eigenvalue decomposition of $G_{\mu\nu}(\bm r)$, this means that the smaller eigenvalue, corresponding to the real space direction with the slowest angle variation, no longer vanishes, $|\lambda_-|>0$. This slowest angle variation now corresponds to a non-coplanar type of deformation of a spiral state. The situation is depicted in Fig.~\ref{fig:vortices}(b) which shows a spiral state deformed such that spins slowly rotate out of the local spiral plane when moving vertically. The rate of this out-of-plane angle variation when moving vertically is given by $\sqrt{|\lambda_-|}$

As long as the eigenvalues $\lambda_+$ and $\lambda_-$ of $G_{\mu\nu}$ are sufficiently different, $|\lambda_+|\gg|\lambda_-|$, and $\sqrt{|\lambda_+|}\approx q_\circ$ the state corresponds to a small deformation of an exact ground state spiral and one can still associate the larger eigenvalue/eigenvector with a momentum field via ${\bm q}=\sqrt{\lambda_+}{\bm v}_+$ (where we assume normalized eigenvectors $|{\bm v}_\pm|=1$). Furthermore, a normal direction ${\bm \Omega}_q$ (with $|{\bm \Omega}_q|=1$) of the spiral can be defined by ${\bm \Omega}_q={\bm S}\times \partial_q {\bm S}$ where $\partial_q$ is a derivative along the wave vector direction $\partial_q={\bm v}_+\cdot{\bm \nabla}$. Since normalized real eigenvectors are generally only defined up to a sign, the signs of ${\bm q}$ and ${\bm \Omega}_q$ are arbitrary. However, as will become very important later,  $\{{\bm q},{\bm \Omega}_q\}$ and $\{-{\bm q},-{\bm \Omega}_q\}$ correspond to identical states.

It follows that small deformations of an exact ground state spiral can be of several types: $(i)$ angle variations of ${\bm q}$, $(ii)$ amplitude variations of $|{\bm q}|$ around $q_\circ$ and $(iii)$ variations in the normal direction ${\bm \Omega}_q$ (which correspond to small non-zero $\lambda_-$). Importantly, the angle variations of ${\bm q}$ in $(i)$ define a U(1) order parameter space associated with the wave vector, in close analogy to the XY case. Additionally, in the Heisenberg case, the spin rotation symmetry allow to perform global spin rotations. Besides rotating the actual spins $\bm S$ such transformations also rotate the normal directions ${\bm \Omega}_q$ of spirals which means that spin rotations must always be applied to the dyad formed by the two orthogonal vectors $\bm S$ and ${\bm \Omega}_q$. This dyad has the rotation properties of a rigid body such that its order-parameter space is SO(3).

We can therefore conclude that, locally, a spin spiral in the Heisenberg model is fully determined by the set of three vectors $\{{\bm q},{\bm S},{\bm \Omega}_q\}$ where $\bm q$ lives in real space and transforms according to U(1) while the last two vectors are defined in spin space and transform according to SO(3). Further taking into account that $\{{\bm q},{\bm S},{\bm \Omega}_q\}$ and $\{-{\bm q},{\bm S},-{\bm \Omega}_q\}$ are identical, the full order parameter space is given by $\text{U(1)}\times\text{SO(3)}/\mathds{Z}_2$.

\subsubsection{Homotopy group analysis}\label{sec:homotopy}

Having identified the system's order parameter space as $\text{U(1)}\times\text{SO(3)}/\mathds{Z}_2$ we can now determine the possible topological defects by calculating the first homotopy group $\pi_1(\text{U(1)}\times\text{SO(3)}/\mathds{Z}_2)$.
Here we discuss this from a more physical perspective, and the rigorous mathematical treatment can be seen in Appendix~\ref{sec:topological_index} and~\ref{sec:homotopy_computation}.

\begin{figure*}[t]
    \centering
    \includegraphics[width = 0.99\linewidth]{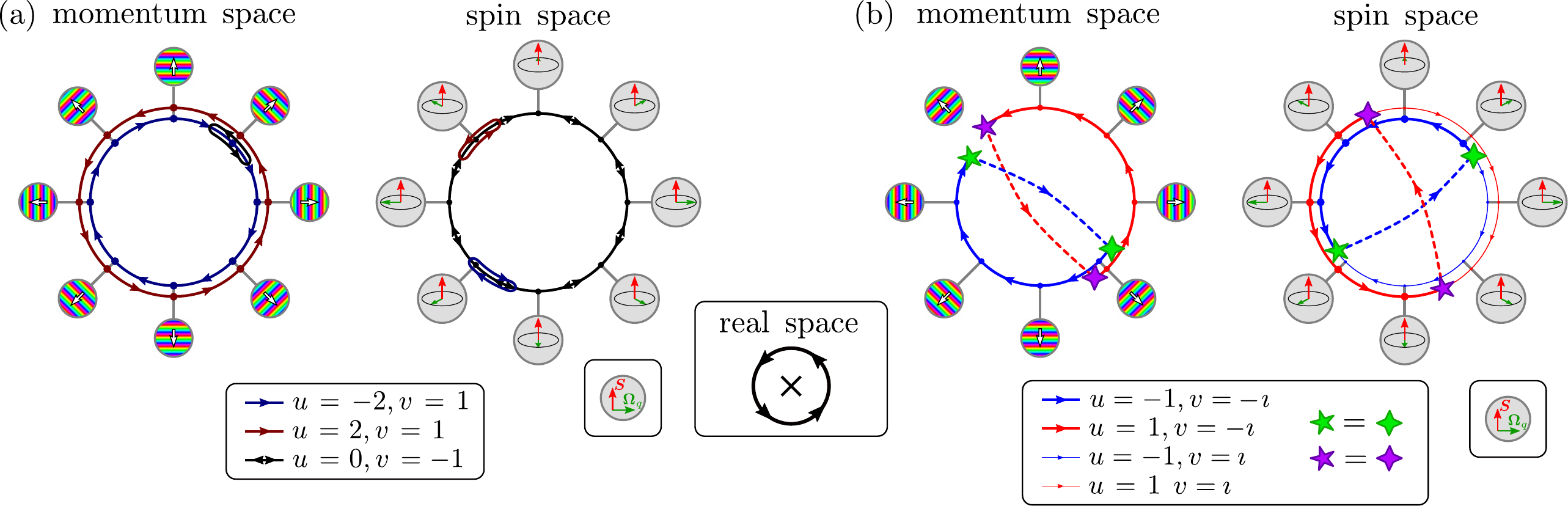}
    \caption{Schematic illustration of vortices with (a) even number of $\pi$ momentum rotations and (b) odd number of $\pi$ momentum rotations. Both subfigures show the change of the spin configuration in momentum space and spin space when traveling around the vortex core (marked by $\times$ in the center inset) along a closed path in counterclockwise direction. Note that the illustrated spin order parameter space (right panel in each subfigure) shows an exemplary $2\pi$ rotation of the dyad $\{{\bm S},{\bm \Omega}_q\}$ about ${\bm S}$. The actual spin order parameter space is SO(3) and the rotation of $\{\bm S,{\bm \Omega}_q\}$ in a vortex can occur about any axis. (a) shows a full momentum antivortex in dark blue with $u=-2$ and no winding in spin space ($v=1$) and an equivalent full momentum vortex with $u=2$ and $v=1$ in dark red. A $\mathds{Z}_2$ vortex with $v=-1$ and no vorticity in momentum ($u=0$) is shown in black, where the direction of the rotation in spin space is irrelevant. (b) illustrates fractional momentum antivortices ($u=-1$) in blue and fractional momentum vortices ($u=1$) in red. Diamond and star symbols (in green and purple) correspond to identical states, enabling a short cut in configuration space. Two topologically inequivalent orientations of the rotations in spin space are shown by thick and thin lines which define the chirality index $v=\pm\imath$.}\label{fig:vortex_schematic}
\end{figure*}

Ignoring for a moment the $\mathds{Z}_2$ redundancy that couples the real space and spin space order parameter subspaces, the topological defects are classified by the known homotopy groups $\pi_1(\text{U(1)})=\mathds{Z}$ and $\pi_1(\text{SO(3)})=\mathds{Z}_2$ such that $\pi_1(\text{U(1)}\times\text{SO(3)})=\mathds{Z}\times\mathds{Z}_2$. Thus, without $\mathds{Z}_2$ redundancy, a vortex is characterized by two independent numbers $\{u',v\}$ where $u'=0,\pm1,\pm2,\pm3,\ldots$ and $v=\pm1$. Here, $u'$ describes the standard vorticity of a planar vector field, in our case the momentum, in analogy to the spin and phase fields in XY ferromagnets and superconducting vortices, respectively. The rotation of momentum around a vortex core is schematically illustrated in Fig.~\ref{fig:vortex_schematic}(a) left panel for $u'=\pm 1$ and is in analogy to the planar vortices in Fig.~\ref{fig:XY-spiral}(c) and (d).   Furthermore, $v=\pm1$ describes the vortex structure of $\{\bm S,{\bm \Omega}_q\}$, where $v=-1$ stands for the presence of a vortex (in the following referred to as $\mathds{Z}_2$ vortex) while $v=1$ corresponds to a trivial state in spin space.

Realizations of $\mathds{Z}_2$ vortices are generally more rarely encountered in physical systems~\cite{Kawamura1984,Kawamura2010,Okubo2010,Rousochatzakis2016,Seabrook2020,Aoyama2020}. A well known example is the classical nearest neighbor Heisenberg antiferromagnet on the triangular lattice~\cite{Kawamura1984} which realizes the planar $120^\circ$ N\'eel order in the ground state. There, the local tripods of three spins with pairwise angles of $120^\circ$ likewise transforms like a rigid body with an SO(3) order parameter space.

Independent of its physical platform, a $\mathds{Z}_2$ vortex has the fundamental property that along a path around the vortex core the local configurations (the dyad $\{{\bm S}, {\bm \Omega}_q\}$ in our spiral system or the tripod in a $120^\circ$ N\'eel state) perform a $2\pi$ rotation about {\it any} axis (where vortices with different rotation axis are topologically equivalent). Crucially, a $\mathds{Z}_2$ vortex is its own antivortex which implies that two $\mathds{Z}_2$ vortices annihilate. The rotation of the dyad $\{{\bm S}, {\bm \Omega}_q\}$ around the vortex core is schematically illustrated in Fig.~\ref{fig:vortex_schematic}(a) right panel. Furthermore, Fig.~\ref{fig:vortices}(c) and (d) show two topologically equivalent real space configurations of $\mathds{Z}_2$ vortices, referred to as type I and type II~\cite{Kawamura1984}, which differ by their rotation axis. While for type I $\mathds{Z}_2$ vortices, the normal direction ${\bm \Omega}_q$ remains fixed and the spin $\bm S$ performs a net $2\pi$ rotation, for type II $\mathds{Z}_2$ vortices, the normal direction ${\bm \Omega}_q$ undergoes a $2\pi$ rotation.  

So far, we have ignored the $\mathds{Z}_2$ redundancy in the order-parameter space. Taking it into account the vortex types already found remain stable topological defects of the system. However, new vortex configurations become possible which use the equivalence of $\{{\bm q},{\bm \Omega}_q\}$ and $\{-{\bm q},-{\bm \Omega}_q\}$ as a ``short cut" through the order parameter space. In an elementary realization of these vortices ${\bm q}$ and ${\bm \Omega}_q$ both perform a $\pi$-rotation when moving once around the vortex core along a closed path such that $\{{\bm q},{\bm S},{\bm \Omega}_q\}$ continuously changes into $\{-{\bm q},{\bm S},-{\bm \Omega}_q\}$. However, since these sets describe the same state, the spin configuration is smooth outside the vortex core. The $\pi$ rotation in ${\bm q}$ implies that this vortex corresponds to a fraction (half) of a $u'=\pm1$ momentum vortex where momentum rotates by $2\pi$. We therefore redefine the topological index via $u\equiv2u'\in\mathds{Z}$ which counts the number of $\pi$ rotations of the momentum when moving around a vortex. This means that elementary half-vortices are described by $u=1$ and $u=-1$ (half-momentum vortex and half-momentum antivortex, respectively) while the vortices in Fig.~\ref{fig:vortex_schematic}(a) have $u=-2,0,2$. The $\pi$ rotations of $\bm q$ and ${\bm \Omega}_q$ are schematically illustrated in Fig.~\ref{fig:vortex_schematic}(b). Furthermore, explicit vortex spin configurations for $u=1$ and $u=-1$ are shown in Fig.~\ref{fig:vortices}(e)-(g).

The preceding discussion demonstrates that vortices in our spiral spin system are divided into two distinct classes, namely fractional vortices for odd $u$ and non-fractional vortices for even $u$. In the case of even $u$, the index $v$ describes the presence or absence of a $\mathds{Z}_2$ vortex. When $u$ is odd, two distinct types of topologically inequivalent spin configurations characterized by an index $v$ also exist. However, this distinction is more subtle and does not correspond to the presence/absence of a vortex but to a hidden structure related to the chirality of the vortex. More precisely, along a closed path around the vortex core, with spin $\bm S$ at the base point, the dyad $\{{\bm S}, {\bm \Omega}_q\}$ performs a net $\pi$ rotation from $\{{\bm S}, {\bm \Omega}_q\}$ to $\{{\bm S}, -{\bm \Omega}_q\}$ where the rotation axis is given by $\bm S$. Regarding $\bm S$ as a screw axis, this rotation can be right-handed (counterclockwise) or left-handed (clockwise). Since these two types of $\pi$ rotations differ by a full $2\pi$ rotation which is topologically non-trivial, the two chiral versions of fractional momentum vortices are topologically distinct. This distinction is schematically illustrated by thin and thick lines in Fig.~\ref{fig:vortex_schematic}(b) right panel. It is also depicted more explicitly in Fig.~\ref{fig:vortices}(f) and (g) showing two chiral versions of half-momentum vortices with $u=1$. As one can see, in Fig.~\ref{fig:vortices}(f) the rotation of ${\bm \Omega}_q$ (green arrow) about the $x$ axis is counterclockwise, while it is clockwise in Fig.~\ref{fig:vortices}(g).

We note that no such topological distinction between different chiralities exists for non-fractional $\mathds{Z}_2$ vortices. This is because a clockwise and a counterclockwise $2\pi$ rotation of $\{{\bm S}, {\bm \Omega}_q\}$ about some axis differ by a $4\pi$ rotation, which is topologically trivial. 

The vorticity $u$ can be computed directly, for both even and odd values of $u$, from the accumulated planar rotation angle along a closed path. On the other hand, the proper definition and calculation of $v$ requires an essential step of abstraction: As explained in Appendix~\ref{sec:topological_index} the SO(3) rotation of $\{{\bm S}, {\bm \Omega}_q\}$ associated with each discrete step along the path around the vortex core is mapped onto an element of SU(2). A $\mathds{Z}_2$ vortex for even $u$ is then identified by the property that an accumulated $2\pi$ rotation along the path corresponds to minus identity in SU(2). As also explained in Appendix~\ref{sec:topological_index}, when $u$ is odd the mapping onto elements of SU(2) naturally leads to a definition of an {\it imaginary} $\mathds{Z}_2$ vorticity $v$ where right-handed (left-handed) vortices carry $v=-\imath$ ($v=\imath$). In summary, the indices $u$ and $v$ have to fulfill the constraint
\begin{align}
u\text{ even }&\leftrightarrow v=\pm1,\notag\\
u\text{ odd }&\leftrightarrow v=\pm\imath.\label{eq:constraint_uv}
\end{align}

\subsubsection{Fusion rules}\label{sec:fusion}
Having identified the possible vortex types in our spiral spin system, it remains to be understood how pairs of them fuse into new vortices or annihilate each other. As discussed in Appendix~\ref{sec:topological_index}, two vortices with $\{u_1,v_1\}$ and $\{u_2,v_2\}$ combine into a new vortex with
\begin{equation}
\{u_1,v_1\}\times\{u_2,v_2\}=\{u_1+u_2,v_1 v_2\},\label{eq:fusion}
\end{equation}
which means that $u$ ($v$) is an additive (multiplicative) index. Note that Eq.~(\ref{eq:fusion}) is consistent with the constraint in Eq.~(\ref{eq:constraint_uv}).

Our results from numerical simulations discussed further below indicate that the physically most relevant vortex types (that involve the smallest excitation energies) are the half-momentum antivortex with $u=-1$ (in the following illustrated by a blue dot $\textcolor{blue}{\bullet}$), the half-momentum vortex with $u=1$ (illustrated by a red dot $\textcolor{red}{\bullet}$) and the $\mathds{Z}_2$ vortex with $u=0$, $v=-1$ (illustrated by a black dot $\bullet$). We note that when $u=\pm 1$ the notation $\textcolor{blue}{\bullet}$/$\textcolor{red}{\bullet}$ does not distinguish between the two chiralities $v=\pm\imath$. Using these conventions, the fusion rules for these three vortex types can the be illustrated as
\begin{align}
\textcolor{blue}{\bullet}\times\textcolor{red}{\bullet}&=1+\bullet\label{eq:fusion1}\\
\textcolor{blue}{\bullet}\times\bullet&=\textcolor{blue}{\bullet} \text{ and } \textcolor{red}{\bullet}\times\bullet= \textcolor{red}{\bullet}\label{eq:fusion2}\\
\bullet\times\bullet&=1\label{eq:fusion3}
\end{align}
The first rule, Eq.~(\ref{eq:fusion1}), states that the $\pi$ spin rotations in $\textcolor{blue}{\bullet}$ and $\textcolor{red}{\bullet}$ may either annihilate and result in the trivial state 1 (if they have opposite chiralities $v_1=-v_2$) or add up to a full $2\pi$ rotation and form a $\mathds{Z}_2$ vortex (if they have the same chiralities $v_1=v_2$). In either case, the $\pi$ momentum rotations of both vortices annihilate each other. The second rule, Eq.~(\ref{eq:fusion2}), states that a $2\pi$ rotation of the dyad $\{{\bm S}, {\bm \Omega}_q\}$ can be added to either $\textcolor{blue}{\bullet}$ or $\textcolor{red}{\bullet}$ without changing its momentum vorticity (however, it changes the chirality $v$). This property is also illustrated in Fig.~\ref{fig:vortex_schematic}(b) where it corresponds to a change of the rotation direction in spin space (right panel) from the thick line to the thin line or vise versa. The third rule, Eq.~(\ref{eq:fusion3}), is a consequence of the $\mathds{Z}_2$ property of $\bullet$ which implies that two $\mathds{Z}_2$ vortices annihilate each other.

Remarkably, when ignoring the distinction between momentum antivortices and vortices $\textcolor{blue}{\bullet}/\textcolor{red}{\bullet}$ these are the same fusion rules as those of the Ising anyons $\sigma$ and $\epsilon$ known from systems with non-Abelian Ising topological order~\cite{Nayak2008}, when identifying \textcolor{blue}{$\bullet$}/\textcolor{red}{$\bullet$} $\leftrightarrow$ $\sigma$ and $\bullet\leftrightarrow\epsilon$. Specifically, these fusion rules read $\sigma\times\sigma=1+\epsilon$, $\sigma\times\epsilon=\sigma$, $\epsilon\times\epsilon=1$. This suggests an analogy between non-Abelian topological order and our spiral spin system. On the other hand, the classical nature of our model is, of course, a fundamental distinction between the two systems.

In principle, an analogous property to Eq.~(\ref{eq:fusion1}) also exists for the fusion of two half-momentum vortices, $\textcolor{red}{\bullet}\times\textcolor{red}{\bullet}=\textcolor{red}{\bullet\bullet}(1+\bullet)$ and analogously for half-momentum antivortices. This follows from the $\mathds{Z}$-valued topological index $u$ of momentum vortices and represents a generalization of the $\mathds{Z}_2$ nature of the $\sigma$ anyon in Ising topological order. As we will see below, however, the resulting full momentum antivortex with $u=2$ is energetically more costly and is rarely found in numerical simulations of the model Hamiltonian.

\subsection{Numerical identification} 

\begin{figure*}[]
    \centering
    \includegraphics[width=2\columnwidth]{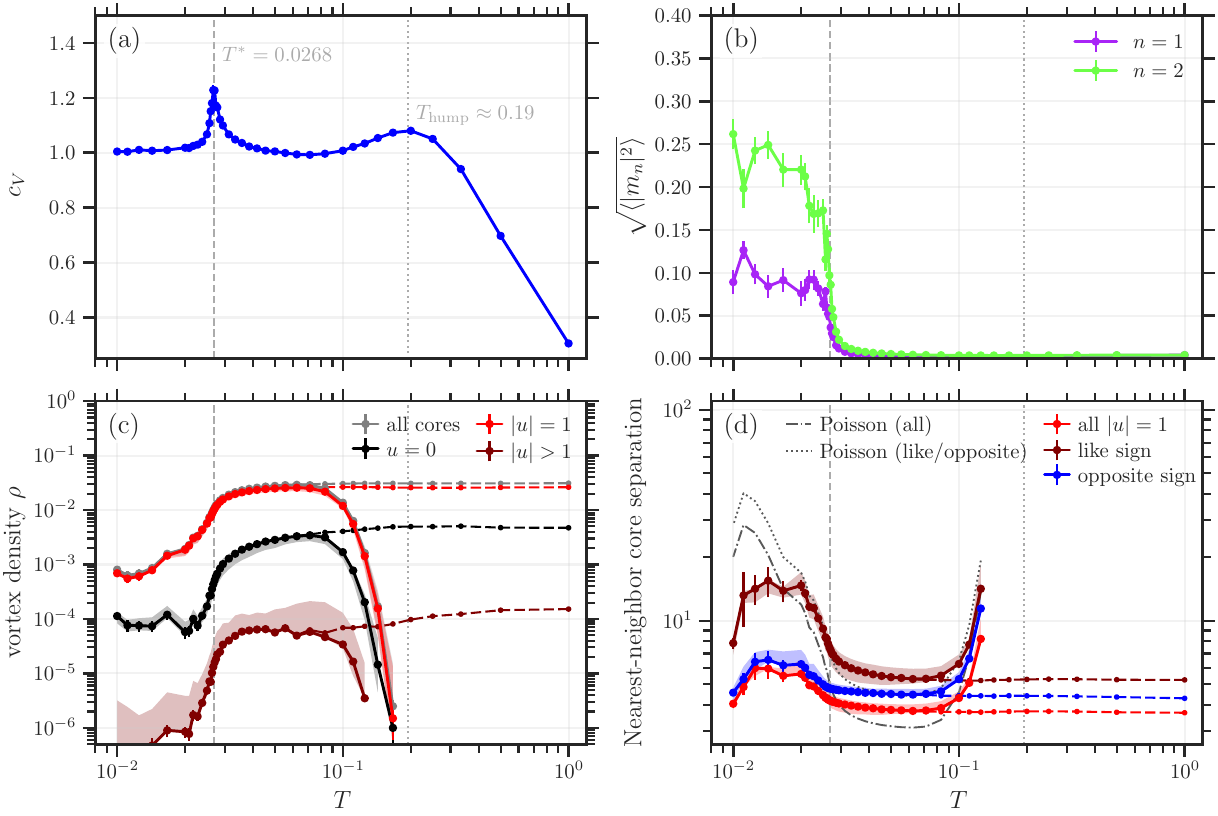}
    \caption{(a) Specific heat capacity as function of temperature. (b) Lattice-nematic order parameters for nearest and second-nearest neighbor bonds. (c) Vortex density as a function of temperature. (d) Mean nearest-neighbor vortex core separation for detector radius $r=3$. Solid lines show the $\langle {\bm q^2} \rangle$ noise-filtered vorticity count, while dashed lines show the result only noise-filtered with loop radius persistence. Shaded bands span the same ($\langle {\bm q^2} \rangle$ noise-filtered) analysis re-run at detector radii $r = 2-4$. The simulations freeze into a metastable configuration with a fixed vortex number below $T=0.02$.}
    \label{fig:vortexdensity}
\end{figure*}

\begin{figure*}[]
    \centering
    \includegraphics[width=2\columnwidth]{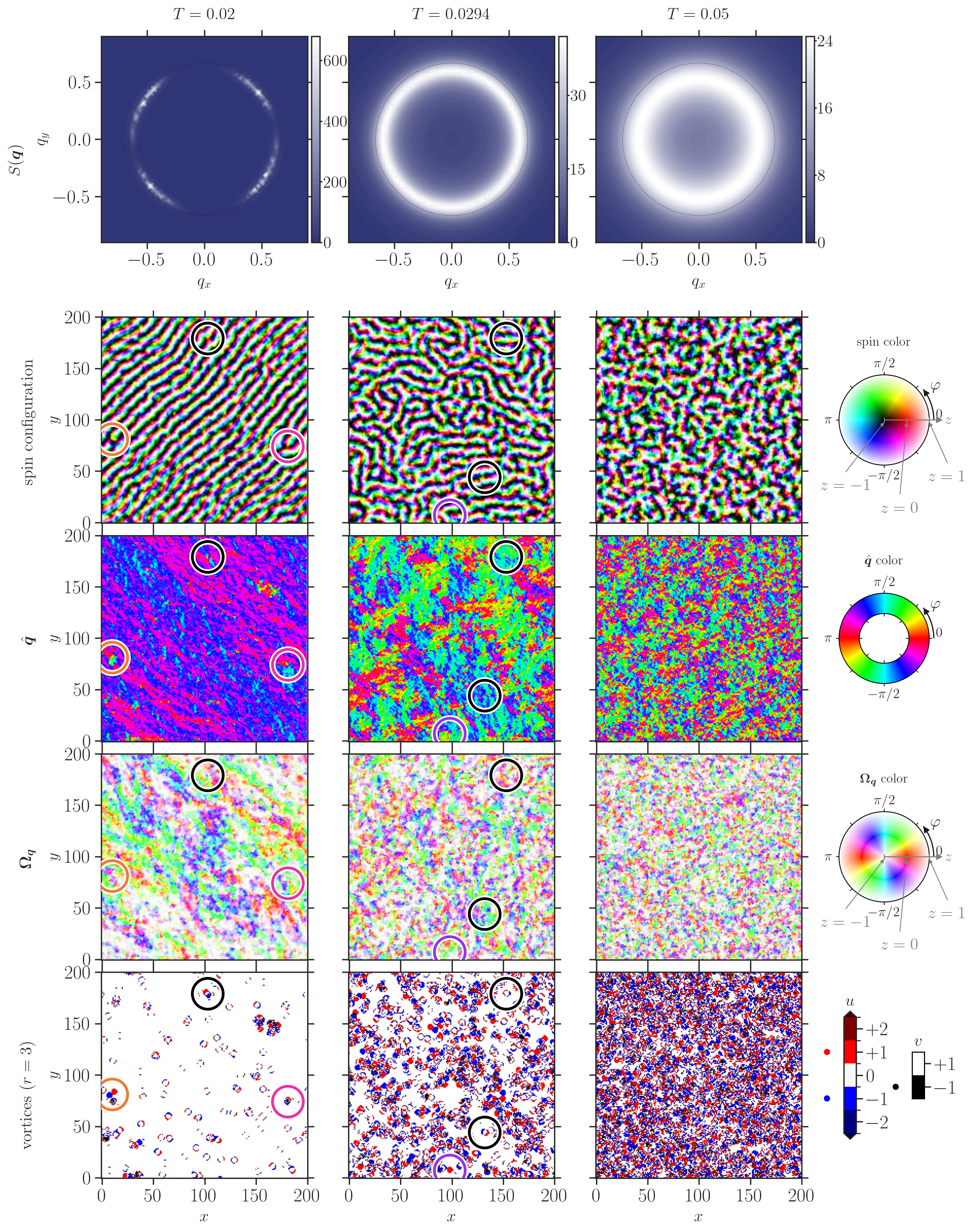}
    \caption{Top row: the spin structure factor (average over all runs) for three representative temperatures $T = 0.02$, $T=0.0294$ and  $T=0.05$. Following rows: one selected spin configuration per temperature with corresponding $\hat{\bm q}$, $\hat{\Omega}_{\hat{\bm q}}$ and vortex configurations ($r=3$). Circles highlight different features from the vortex configurations: (black) three tightly bound $|u|=1$ vortex-antivortex pairs that fuse to the vacuum (one at the lowest temperature and two at the middle temperature), (pink) one tightly bound $|u|=1$ vortex-antivortex pair that fuses to a $\mathds{Z}_2$ vortex (at the lowest temperature), (purple) one isolated $u=+1$ vortex (at the middle temperature) and (orange) one loosely bound $|u|=1$ vortex-antivortex pair (at the lowest temperature).}
    \label{fig:spinconfig}
\end{figure*}

Next, we demonstrate how the analytically derived vortex types manifest as finite-temperature excitations in Heisenberg spiral spin liquids using classical MC simulations (see Methods, Section~\ref{sec:methods}, for details on the simulations). We focus on a $200\times200$ square lattice with periodic boundary conditions. We set $\delta = 0.03$, giving a small, nearly circular ring of radius $q_\circ = 0.66$, suppressing order-by-disorder effects down to low temperatures.

In Fig.~\ref{fig:vortexdensity}(a), we show the specific heat capacity as a function of temperature $T$, showing a broad hump around $T_{\rm hump}\approx 0.19 \pm 0.04$ and a sharp peak at $T^*\approx 0.0268 \pm 0.0004$. Thermal average spin structure factors 
\begin{equation}
S(\bm q) = \frac{1}{N}\sum_{i,j}\bigl\langle \bm S_i\cdot\bm S_j\bigr\rangle \;e^{i\bm q\cdot(\bm r_i-\bm r_j)}
\end{equation}
and representative real-space spin configurations are shown in Fig.~\ref{fig:spinconfig}. Above $T\approx0.2$ (not shown), the system is in an uncorrelated paramagnetic regime, and the broad feature at $T_{\rm hump}$ marks the onset of spiral correlations which have wave vectors lying on or inside the ring-minimum. Just below $T_{\rm hump}$, the system enters a so-called pancake liquid regime~\cite{Shimokawa2019Ripple}, in which the spin structure factor has strong weight throughout the entire area enclosed by the spiral ring. The pancake liquid continuously evolves into a ring liquid (see $T = 0.05$ and $T=0.0294$) upon cooling. Below the sharp peak at $T^*$, the spin structure breaks lattice-rotational symmetry, as quantified by the lattice-nematic order parameter
\begin{equation}
m_n = \frac{1}{2N}\sum_i\sum_{j\in n} e^{2i\theta_{ij}} {\bm S}_i\cdot{\bm S}_j,
\end{equation}
where $\theta_{ij}$ is the orientation of the bond $ij$ and the sum over $j$ runs over the $n$th neighbor-directions.
This lattice-nematic order parameter is shown as a function of temperature in Fig.~\ref{fig:vortexdensity}(b), making it clear that the system enters a lattice-nematic phase below $T^*$ (see also spin configuration in Fig.~\ref{fig:spinconfig} for $T=0.02$). We thus identify the sharp peak at $T^*$ as a phase transition between a disordered spiral spin liquid and an ordered lattice-nematic phase.

Linear spin-wave calculations show that along the ring-minimum, single-${\bm q}$ states with ordering wave vectors along diagonals are entropically favored by spin waves at $\delta = 0.03$, see Appendix~\ref{sec:lsw_entropy}. This is likely the explanation for the low-$T$ phase showing broad peaks centered on the diagonals in the well-equilibrated regime (see $T=0.02$ in Fig.~\ref{fig:spinconfig}).

The ideal vortex configurations shown in Fig.~\ref{fig:vortices} are difficult to identify by eye in the real-space spin configurations obtained from MC simulations in Fig.~\ref{fig:spinconfig}. This is due to thermal fluctuations and the short length scales of phase coherence associated with the spin spirals. Therefore, a more reliable way to identify vortices is to calculate the vorticities $u$ and $v$, which are shown in the last row of Fig.~\ref{fig:spinconfig}. As described in Methods, Section~\ref{sec:methods}, $u$ and $v$ are calculated for loops of finite radius $r$. Choosing a radius of several lattice spacings ($r=3$ in Fig.~\ref{fig:spinconfig}) makes the vortices more readily visible, as they appear as disks of the same radius. This representation also highlights the fusion of nearby vortex pairs (idealized versions of these disks are shown next to the color bars in the bottom row of Fig.~\ref{fig:spinconfig}). We note that the bottom row of Fig.~\ref{fig:spinconfig} shows the unprocessed raw data for $u$ and $v$ obtained with $r=3$, which become increasingly noisy with increasing temperature. This is because, in addition to well-defined vortices that appear as deformations of phase-coherent spin spirals, nonzero vorticity can also arise from completely random spin configurations when a given loop happens to enclose a finite vorticity by chance. Below, we discuss how these accidental vortex configurations can be filtered out. Nevertheless, the raw data already reveal characteristic vortex configurations, highlighted by circles: (i) three tightly bound $|u|=1$ vortex-antivortex pairs that fuse to the vacuum (one at the lowest temperature and two at the middle temperature), (ii) one tightly bound $|u|=1$ vortex-antivortex pair that fuses to a $\mathds{Z}_2$ vortex (at the lowest temperature), (iii) one isolated $u=+1$ vortex (at the middle temperature), and (iv) one loosely bound $u=\pm 1$ vortex-antivortex pair (at the lowest temperature).

Generally, the plots show that within the spin liquid regime, there is a dense configuration of vortices mainly of $u=\pm 1$, which becomes more dilute as temperature decreases. The highlighted circles (i) in the middle panel indicate that the dilution comes from tightly bound $|u|=1$ vortex-antivortex pairs that fuse to the vacuum. The tightly bound pairs of type (i) also seems to be the most common case within wide temperature ranges, being visible in all three panels. The tightly bound $|u|=1$ vortex-antivortex pairs that fuse to a $\mathds{Z}_2$ vortex (ii) are considerably rarer than (i). This is because a $\mathds{Z}_2$ vortex contains a point of singularity associated with a high energy cost. Interestingly, pure isolated $\mathds{Z}_2$ vortices that do not appear as bound $|u|=1$ vortex-antivortex pairs are not not as common in our simulation results. The isolated $|u|=1$ vortices (iii) are also only visible at higher temperatures due to their energy cost. The loosely bound $u=\pm 1$ vortex-antivortex pair (iv) is common at the intermediate temperatures but increasingly rare at low temperatures.

To further characterize the behavior as a function of temperature, we compute the total vortex density $\rho$. For a reliable vortex count, vortices that are not associated with well-defined spin spirals but instead arise from finite vorticity occurring by chance in random paramagnetic spin configurations must be filtered out. As described in more detail in Methods, Section~\ref{sec:methods}, this is achieved by two separate measures: First, the vorticity must persist at a second radius. Second, the average momentum amplitude $\langle q^2 \rangle$ in an area around the vortex must not significantly exceed the ring minimum $q_\circ^2$, i.e. the vortex is located in a correlated region. The result is shown in Fig.~\ref{fig:vortexdensity}(c), where the vortex density is illustrated for each vortex type. For comparison we also show the vortex densities without the $\langle q^2 \rangle$ filter by dashed lines. When applied to random paramagnetic spin configurations, the detector reports a vortex density of $\sim3\%$ without the $\langle q^2 \rangle$ filter, which is consistent with the high-$T$ MC results. As the system enters the ring-liquid regime upon cooling, the correlated spiral vortices start prevailing over thermal noise, which is supported by the unfiltered curve agreeing with the noise-filtered one. Furthermore, in this regime ($T=0.05$ in Fig.~\ref{fig:spinconfig}) the raw vorticities contain clear disk-like patterns, showing that one can start to separate real spiral vortices from a background of noise.

Looking at Fig.~\ref{fig:vortexdensity}(c), it is clear that in the paramagnetic regime, the vorticity is only thermal, and no correlated vortices are found. Then, with the onset of spiral correlations at $T_{\rm hump}$, the vortex density starts to rapidly increase, until around $T=0.07$, where the system reaches a clear ring liquid regime and a vortex density around $3\%$. The vortex density stays approximately constant at $3\%$ within the ring-liquid regime, with most vortices being ideitified as $|u|=1$ vortices. The $u=0$ $\mathds{Z}_2$ vortices have a density of about $0.3\%$, and vortices with $|u|>1$ remain extremely rare even in this regime. 

As the system crosses from the ring-liquid into the lattice-nematic phase, the vortex density rapidly decreases, which can be associated with the $|u|=1$ vortex-antivortex pairs getting so tightly bound that many of them fuse to vacuum. Remarkably, a small but finite vortex density is still observed in the ordered regime, despite a background of mostly constant momentum $\bm q$. As shown in the fourth row of Fig.~\ref{fig:spinconfig}, this is because $\bm \Omega_{\bm q}$ still shows strong spatial fluctuations at the lowest temperatures without any preferred orientations (we note that the color scale for $\bm \Omega_{\bm q}$ takes into account the redundancy between $\bm \Omega_{\bm q}$ and $-\bm \Omega_{\bm q}$ by plotting both with the same color). As the temperature is further lowered, the vortex density keeps decreasing until the MC simulations freeze below $T=0.02$ due to the vortex dynamics slowing down.

We also compute the mean nearest-neighbor separation between vortex cores as function of temperature in Fig.~\ref{fig:vortexdensity}(d). Comparing this to the expected separation of uncorrelated vortices (Poisson distributed) $1/(2\sqrt{\rho})$, it is apparent that in the spiral-spin-liquid regime, the vortices are slightly repelling, whereas in the low-temperature ordered regime they are attracting. If we separate $|u|=1$ pairs into pairs of like and opposite sign, we find that the like sign pairs are repelling in the spiral-spin-liquid regime and Poisson distributed in the (equilibrated) low-temperature regime. For the opposite sign pairs, they are closer to Poisson distributed in the spiral-spin-liquid regime, while they form tightly-bound vortex-antivortex pairs in the low-temperature regime. This interpretation is also supported by our earlier observations in the vortex configurations of Fig.~\ref{fig:spinconfig}. Overall, this substantiates the physical picture in which free fractional vortices dominate the system's thermal fluctuations at intermediate temperatures, before becoming closely bound and dilute as the system enters the ordered phase.

\section{Discussion}\label{sec:discussion}

Concepts such as fractionalization and topological order have attracted immense interest in condensed matter physics. A particularly rich class of systems in which these concepts emerge are quantum spin liquids, where fractionalized quasiparticles are a defining feature of this phase. Quantum spin liquids can host several forms of topological order, including $\mathds{Z}_2$ topological order, Ising topological order, and chiral topological order in Laughlin-like analogues of fractional quantum Hall states. These forms of topological order are inherently quantum phenomena, whose realization requires quantum effects.

Here, we demonstrate that complicated fractional topological defects with fusion rules analogous to those of Ising topological order in certain aspects can also arise in a purely classical spin system: a classical spiral spin liquid with spin-isotropic Heisenberg interactions. The fractionalization of these defects originates from a $\pi$ winding of the spiral wave vector around the vortex core, accompanied by a reversal of the normal direction of the spiral plane. Pairs of opposite fractional vortices can fuse either into a topologically trivial state or into a $\mathds{Z}_2$ vortex. Our numerical simulations identify these unusual vortex excitations over a broad range of intermediate and low temperatures. Although we focus on the spiral spin liquid realized in the $J_1$-$J_2$-$J_3$ square-lattice Heisenberg model at a particular $\delta=0.03$, the underlying mechanism is more general and is expected to apply to wider ranges of $\delta$ as well as to spiral spin liquids on other two-dimensional lattices, including the $J_1$-$J_2$-$J_3$ triangular~\cite{Glittum2021, Liu2022, Glittum2026} and $J_1$-$J_2$ honeycomb models~\cite{Shimokawa2019MultipleQ}. The key requirement is the existence of a spiral contour which should be approximately circular, thereby suppressing order-by-disorder mechanisms that would otherwise select specific wave vectors. Remarkably, our simulations further show that the fractional vortices possess a certain degree of immunity to magnetic ordering: a small concentration of tightly bound fractional vortex-antivortex pairs persists even below the order-by-disorder transition. This robustness also suggests that fractional vortices should survive weak deviations from the idealized parameter regime with an exactly degenerate spiral contour, where weak perturbations instead select a set of preferred wave vectors.

Our results have direct implications for magnetic materials. In particular, they suggest that fractional momentum vortices should occur whenever two relatively mild conditions are satisfied: $(i)$ the spin structure factor exhibits a ring-like pattern of magnetic scattering, and $(ii)$ the spin interactions are of Heisenberg type. Condition $(i)$, however, need not be strictly fulfilled, as our results show that fractional vortices can persist in the presence of weak magnetic order. Condition $(ii)$ is likewise not particularly restrictive, since Heisenberg interactions provide a standard description of many magnetic systems. A possible additional requirement is that quantum fluctuations should not be strong enough to destroy well-defined local spin moments. Importantly, the observation of fractional vortices at intermediate temperatures relaxes this constraint: it is sufficient that thermal fluctuations dominate over quantum fluctuations within the relevant intermediate temperature regime.

Given the large number of materials realizing spiral spin liquid phases, we expect our discovery to be readily applicable to experiments.
For example, our results are directly relevant to known two-dimensional Heisenberg spiral spin liquid materials, such as the triangular delafossite-like compound AgCrSe$_2$~\cite{Andriushin2025} and honeycomb FeCl$_3$~\cite{Gao2022}, where spiral-ring features have been observed in neutron-scattering experiments. A particularly compelling direction would be the direct magnetic imaging of individual fractional vortices, which our results suggest should occur in these materials. Although resolving these objects requires challenging atomic-scale spatial resolution, rapid experimental progress towards real-space magnetic imaging~\cite{Zhao2019} makes the direct observation of fractional vortices an increasingly realistic prospect.
%Our results hold in general for spin-isotropic Heisenberg models with a ring-minimum in the momentum-space exchange coupling. In addition to the $J_1$-$J_2$-$J_3$ model on the square lattice, this includes also the $J_1$-$J_2$-$J_3$ triangular and $J_1$-$J_2$ honeycomb models. In particular, such Heisenberg spiral spin liquids have already been observed experimentally for the triangular delafossite-like AgCrSe$_2$~\cite{Andriushin2025} and honeycomb FeCl$_3$~\cite{Gao2022}.

\section{Methods}\label{sec:methods}

\subsection{Monte Carlo simulations}

The MC simulations are done on periodic square lattices of linear system size $L=200$, corresponding to $N=200^2=4\times10^4$ spins in total. Each temperature run consists of $10^6$ MC steps, and each MC step consists of $N$ repetitions of a single-spin Metropolis update followed by $5$ over-relaxation reflections about the local exchange field. For all Metropolis updates the proposed spin is drawn from an adaptive Gaussian cone whose width is adjusted during equilibration to approach a $50\%$ acceptance rate~\cite{Alzate-Cardona2019}.

Every run is equilibrated by adiabatic cooling, in which $\beta$ is ramped from $\beta = 0.1$ to its target value before sampling begins. Because the autocorrelation time varies by three orders of magnitude across the temperature range studied, the two halves of the temperature range were run with different schedules and lengths.

For $\beta \le 32$ we use a linear ramp in $\beta$ with $9\times10^4$ MC steps down to the target $\beta$ and $10^4$ MC steps at the target $\beta$. 

For $\beta \ge 34$, slowing down sets in, so we use $5\times10^6$ equilibration steps. The ramp occupies $90\%$ of the equilibration steps, and is piecewise-linear in $\beta$ and concentrated on the transition region: First, we do a relatively fast approach of $10^5$ steps from $\beta = 0.1$ to $\beta = 34$, then a slow cooling within the window $\beta \in [34, 44]$. For target temperatures inside that window, the entire remaining $4.4\times10^6$ steps are spent reaching the target. For colder targets the window is crossed in $2.6\times10^6$ steps and the remaining $1.8\times10^6$ complete the cooling down to the target $\beta$. The final $5\times10^5$ equilibration steps are performed at the target $\beta$.

We perform 10 statistically independent runs at each of the $\beta$ values, and all quoted data and uncertainties are the average and standard error over these runs.

Energy, magnetization and the bond-nematic order parameters are accumulated every MC step, the structure factor $S(\bm{q})$ is measured every 100 MC steps, and the full spin configuration is stored every $10^4$ MC steps, giving a total of 100 configurations per run to be used in the vortex detection.

The specific heat is calculated from the energy fluctuations of each run, $c_V = \beta^2 N\left(\langle e^2\rangle - \langle e\rangle^2\right)$, where $\langle e \rangle$ is the average energy per site.

\subsection{Vortex-detection in MC configurations}

For each spin configuration (1000 configurations per $\beta$), $G_{\mu\nu}({\bm r})$ is constructed and diagonalized to give $\bm{q}$ and ${\bm \Omega}_{\bm q}$, from which vortices can be identified. 

The $u$ and $v$ winding numbers are computed on lattice loops of radius $r = 3$. The winding of the sign-free doubled director $\exp(2i\theta_{\hat {\bm q}})$ of the normalized momentum field $\hat {\bm q}$ gives the integer momentum index $u$, where the doubling is introduced to account for ${\bm q} \to -{\bm q}$. The winding of the dyad $\{{\bm S},{\bm \Omega}_q\}$ mapped onto the SU(2) matrix $T$ gives $v$, as described in Appendix~\ref{sec:topological_index}. This winding is also path-combed along the loop, i.e. ${\bm \Omega_{\bm q}}$ is flipped at each step to agree with the previous one, so that no global branch cut is introduced from $\{{\bm q}, {\bm S},{\bm \Omega}_{\bm q}\} = \{-{\bm q}, {\bm S},-{\bm \Omega}_{\bm q}\}$.

Each vortex candidate must persist at a second radius ($r+1$ for $\mathds{Z}_2$ disks, $r-1$ for (half-)momentum vortices). This is done to separate between topological vortices and thermal noise in the correlated regimes. However, in the paramagnetic regime, it is not sufficient as thermal noise is dominating and the probability of detecting a vortex in a paramagnetic state is $3\%$, independent of loop radius. We thus add an additional requirement that the averaged ${\bm q^2}$ on a $15\times15$ site patch surrounding the vortex cores fulfills $\langle {\bm q}^2 \rangle < 0.5$ (motivated by $q_\circ^2 = 0.44$ for the ring-minimum), i.e. the vortex core is located in a correlated region.

Uncertainties on all core counts are obtained by averaging within each run first and then bootstrapping over runs (2000 resamples), because configurations within a run are not necessarily independent.

\section{Acknowledgements}
C.G. acknowledges funding from the European Union's Horizon Europe research and innovation programme under the Marie Sk\l{}odowska-Curie Grant Agreement No. 101126636. 
H.Y. acknowledges the Grant-in-Aid for Research Activity Start-up (Grant No. JP24K22856) and Grant-in-Aid for Early-Career Scientists (Grant No. JP26K17090)  from the Japan Society
for the Promotion of Science.
J.R. acknowledges support from the Deutsche Forschungsgemeinschaft (DFG, German Research Foundation), within Project-ID 277101999 CRC 183 (Project A04).
The computations were performed on resources provided by Sigma2 - the National Infrastructure for High Performance Computing and Data Storage in Norway.

\bibliography{ref}

\appendix
\section{Definition of the topological index $v$}\label{sec:topological_index}
Here, we discuss the definition and calculation of the $\mathds{Z}_2$ topological index $v$ for the rotation of $\{\bm S,{\bm \Omega}_q\}$ when moving along a closed counterclockwise path around the vortex core. The considered closed path consists of $N$ sites where we identify $N+1\equiv1$. The spin configuration at each site is characterized by the dyad $\{\bm S^i,{\bm \Omega}^i_q\}$ where the superscript indicates the site $i\in\{1,2,\ldots, N\}$. One can now uniquely find SO(3) rotation matrices $t_{{\bm m}^i}(\alpha^i)$ about the axes ${\bm m}^i$ (with $|{\bm m}^i|=1$) and angles $\alpha^i\in(-\pi,\pi]$ such that
\begin{equation}
{\bm S}^{i+1}=t_{{\bm m}^i}(\alpha^i){\bm S}^i \text{ and } {\bm \Omega}_{\bm q}^{i+1}=t_{{\bm m}^i}(\alpha_i){\bm \Omega}_{\bm q}^i.
\end{equation}

We start the discussion with the case of even $u$ which means that the momentum vector performs an even multiple of $\pi$ rotations along this path. For a full motion along the closed path, the spin configuration comes back to itself which implies
\begin{equation}
t_{\text{tot}}=t_{{\bm m}^N}(\alpha^N)t_{{\bm m}^{N-1}}(\alpha^{N-1})\cdots t_{{\bm m}^1}(\alpha^1)=\mathds{1}_{3\times3},\label{eq:ttotso3}
\end{equation}
with the three-dimensional identity matrix $\mathds{1}_{3\times3}$. This product does not provide any distinguishing property between the trivial state and a $\mathds{Z}_2$ vortex.

On the other hand, the identification of a $\mathds{Z}_2$ vortex becomes possible when mapping the SO(3) matrices $t_{{\bm m}^i}(\alpha^i)$ onto $2\times2$ SU(2) matrices $T_{{\bm m}^i}(\alpha^i)$ using $T_{{\bm m}^i}(\alpha^i)=e^{-\imath \alpha^i{\bm m}^i\cdot{\bm \sigma}/2}$ where ${\bm \sigma}$ are the Pauli matrices. For a closed path around a $\mathds{Z}_2$ vortex it follows
\begin{equation}
T_\text{tot}=T_{{\bm m}^N}(\alpha^N)T_{{\bm m}^{N-1}}(\alpha^{N-1})\cdots T_{{\bm m}^1}(\alpha^1)=-\mathds{1}_{2\times2},\label{eq:ttot}
\end{equation}
since the $2\pi$ rotation of the dyad $\{\bm S,{\bm \Omega}_q\}$ generates the phase $e^{\imath\pi}=-1$ in SU(2), which describes the double cover of SU(2) by SO(3). On the other hand if $\{\bm S,{\bm \Omega}_q\}$ performs no net rotation (or a rotation with an angle that is an even multiple of $2\pi$) one finds $T_\text{tot}=\mathds{1}_{2\times2}$. This allows one to define the index $v$ via
\begin{equation}
T_\text{tot}=v\mathds{1}_{2\times2},
\end{equation}
with $v=\pm1$, indicating the presence or absence of a $\mathds{Z}_2$ vortex.

Next, we discuss the definition of the topological index $v$ when $u$ is odd. In this case the dyad $\{{\bm S},{\bm \Omega}_q\}$ at the beginning of the closed path changes into $\{{\bm S},-{\bm \Omega}_q\}$ at its end (which, due to the simultaneous $\pi$ rotation of the momentum, describe the same state). Consequently, the change of $\{{\bm S},{\bm \Omega}_q\}$ corresponds to a $\pi$ rotation of ${\bm \Omega}_q$ about the axis $\bm S$, which is the spin direction at the base point of the path. The total product of SO(3) rotations for all steps along the path, $t_\text{tot}$, as defined in Eq.~(\ref{eq:ttotso3}) gives a unique $3\times3$ matrix that does not enable a distinction between different vortex types. For example, for a spin in $z$ direction the $\pi$ rotation yields $t_\text{tot}=\text{diag}(1,-1,-1)$. Again a distinction become possible when mapping the SO(3) rotations onto SU(2) matrices and calculating the product $T_\text{tot}$ as given in Eq.~(\ref{eq:ttot}). Specifically, the promotion to SU(2) allows the distinction between counterclockwise ($+\pi$, right-handed) and clockwise ($-\pi$, left-handed) rotations of ${\bm \Omega}_q$ about the axis ${\bm m}={\bm S}$, since they are different elements in SU(2). Explicitly,
\begin{equation}
T_\text{tot}=e^{\mp\imath\pi{\bm S}\cdot{\bm \sigma}/2}=\mp\imath {\bm S}\cdot{\bm \sigma},\label{eq:ssigma}
\end{equation}
where the upper (lower) sign corresponds to a $+\pi$ ($-\pi$) rotation. We use this result to define the chirality index $v$ via
\begin{equation}
T_\text{tot}=v {\bm S}\cdot{\bm \sigma},\label{eq:defv}
\end{equation}
with $v=\pm\imath$.

For this definition to be meaningful we have to show that $(i)$ the index $v$ does not depend on the choice of where to put the base point on a given path (i.e., $v$ does not depend on the spin $\bm S$ at the base point), $(ii)$ for a fixed base point the index $v$ does not change upon deforming the path, as long as it encloses the vortex, and $(iii)$ the definition should yield fusion rules which specify the index $v$ resulting from the combination of two vortices characterized by $v_1$ and $v_2$. 
\begin{figure}[t]
    \centering
    \includegraphics[width = 0.8\linewidth]{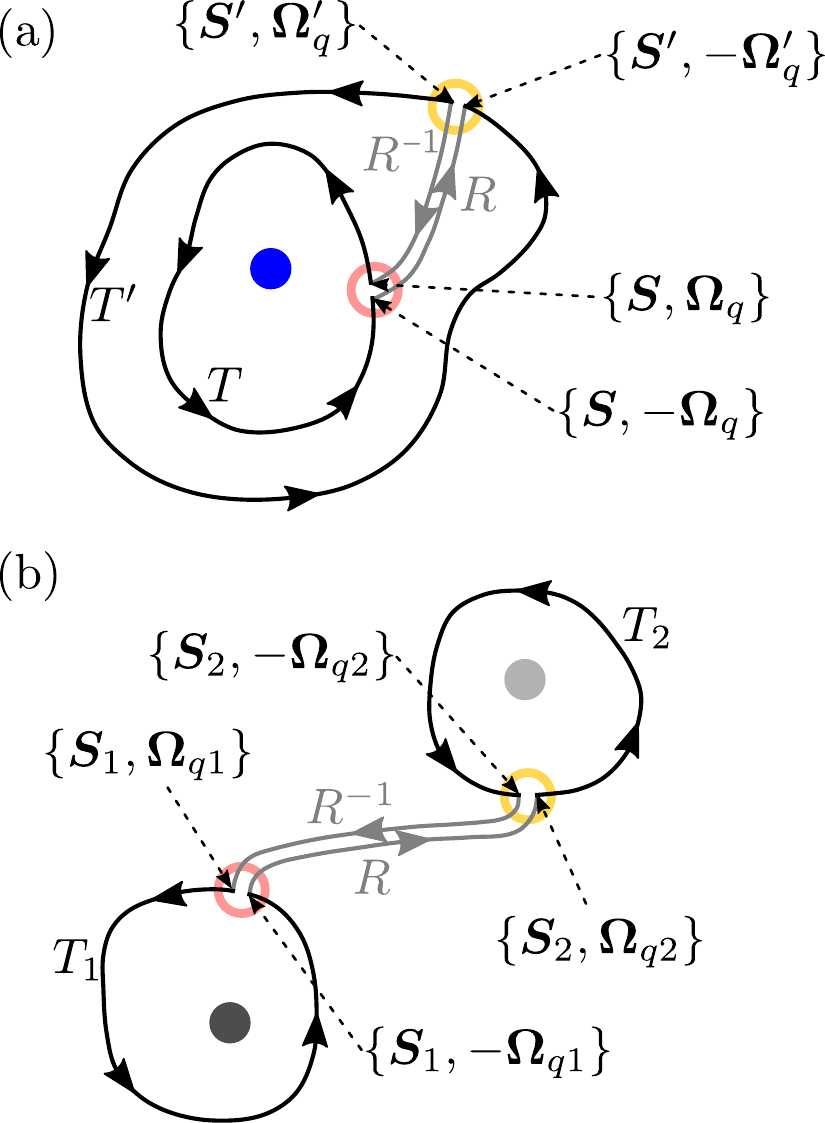}
    \caption{Schematic illustration of vortices (blue and gray disks) surrounded by closed paths to demonstrate the unique definition of $v$ in (a) and to derive the fusion rules for vortices in (b). See text for details. The oppositely oriented gray paths $R$ and $R^{-1}$ coincide spatially but are shown with an offset to improve visual clarity. Light red and yellow circles indicate base points of paths where the spin configurations charcaterized by the dyads $\{{\bm S},{\bm \Omega_q}\}$ are given. At a given point in space, the normal direction ${\bm \Omega}_q$ can have different signs for different locations along the paths, as indicated by the dyads.}\label{fig:definition_v}
\end{figure}

We treat the points $(i)$ and $(ii)$ together considering the configuration of paths shown in Fig.~\ref{fig:definition_v}(a). There, a vortex (blue dot) is surrounded by two arbitrary counterclockwise oriented paths labeled $T$ and $T'$ with different spins $\bm S$ and ${\bm S}'$ at their base points. We will show that the indices $v$ and $v'$ associated with both paths are identical.

In Fig.~\ref{fig:definition_v}(a), the base point of path $T$ ($T'$) is marked by a light red (yellow) circle, where the local spiral is characterized by $\{{\bm S},\pm{\bm \Omega}_q\}$ ($\{{\bm S}',\pm{\bm \Omega}'_q\}$). Here, the different signs $\pm$ indicates that at the same position in space but at different locations along the path (start or end) ${\bm \Omega}_q$ can differ by a sign. Furthermore, the base points of $T$ and $T'$ are connected by an open path $R$. We use the convention of Eqs.~(\ref{eq:ttotso3}) and (\ref{eq:ttot}) and denote the total accumulated SO(3) rotation matrices along these paths by the lower case letters $t$, $t'$ and $r$ and the corresponding total accumulated SU(2) rotation matrices by the upper case letters $T$, $T'$ and $R$ (where we omit the label ``$\text{tot}$''). More specifically,
\begin{equation}
t{\bm S}={\bm S},\quad t{\bm \Omega}_q=-{\bm \Omega}_q,
\end{equation}
and identical equations for $t'$, ${\bm S}'$ and ${\bm \Omega}'_q$. Furthermore,
\begin{equation}
r{\bm S}={\bm S}',\quad r{\bm \Omega}_q={\bm \Omega}'_q,\label{eq:rs}
\end{equation}
and from Eq.~(\ref{eq:ssigma}) and the definition in Eq.~(\ref{eq:defv}) it follows
\begin{equation}
T=v{\bm S}\cdot{\bm \sigma},\quad T'=v'{\bm S'}\cdot{\bm \sigma}.\label{eq:vvp}
\end{equation}
Next, we exploit the property that the area between the paths does not contain a vortex, so the accumulated SU(2) rotation gives the trivial element. Specifically, starting at the base point of $T$, then moving along $T\rightarrow R\rightarrow T'^{-1}\rightarrow R^{-1}$ and coming back to the base point of $T$ gives
\begin{equation}
R^{-1}T'^{-1}RT=\mathds{1}_{2\times2}\;\Longleftrightarrow\;T'=RTR^{-1},\label{eq:rtr}
\end{equation}
where the ordering of the matrices is reversed relative to the order in which the path is traveled, see also Eq.~(\ref{eq:ttot}). We note in passing that in the case of non-fractional vortices with $T=\pm \mathds{1}_{2\times2}$ the matrix $R$ that describes the connection between both paths would drop out of the equation. In our case, however, $R$ plays an essential role in connecting both paths and must be kept. Inserting the first equation of Eq.~(\ref{eq:vvp}) into the second equation of Eq.~(\ref{eq:rtr}) yields
\begin{equation}
T'=v{\bm S}\cdot (R{\bm \sigma} R^{-1})=v{\bm S}\cdot(r^{-1}{\bm \sigma})=v(r{\bm S})\cdot{\bm \sigma}.
\end{equation}
In the second step, we have exploited the fundamental property of the group homomorphism between SO(3) and SU(2), $R{\bm \sigma} R^{-1}=r^{-1}{\bm \sigma}$, and in the third step we have used $r^\text{T}=r^{-1}$. Inserting Eq.~(\ref{eq:rs}) on the right hand side and comparing with Eq.~(\ref{eq:vvp}) we indeed find $v=v'$.

Finally, addressing point $(iii)$, we derive the fusion rules for $v$ for any two (fractional or non-fractional) vortices, considering the configuration in Fig.~\ref{fig:definition_v}(b). It consists of two vortices (light gray and dark gray), surrounded by the counterclockwise paths $T_1$ and $T_2$ which are connected by the open path $R$. The spins at the base points are ${\bm S}_1$ and ${\bm S}_2$, respectively.

To find the fusion rules, we traverse a path around both vortices. Without loss of generality, we start at the base point of the first vortex and then go along the path $T_1\rightarrow R\rightarrow T_2\rightarrow R^{-1}$ and come back to the base point. The total SU(2) rotation along this path, denoted $T$, is then given by
\begin{equation}
T=R^{-1}T_2 RT_1.\label{eq:two_paths}
\end{equation}
Note again the reversed order of the matrix product. Depending on whether the fused vortex is fractional or not, $T$ is given by
\begin{equation}
T=v{\bm S}_1\cdot{\bm \sigma}\text{ or }T=v\mathds{1}_{2\times2},\label{eq:t}
\end{equation}
respectively. If the second vortex is non-fractional, i.e., $T_2=v_2\mathds{1}_{2\times2}$, Eq.~(\ref{eq:two_paths}) yields $T=v_2T_1$. In both cases, when the first vortex is non-fractional ($T_1=v_1\mathds{1}_{2\times2}$) or fractional ($T_1=v_1{\bm S}_1\cdot{\bm \sigma}$) we find $v=v_1 v_2$ by comparing with Eq.~(\ref{eq:t}).

The discussion so far covers the cases where at least one vortex is non-fractional. The fusion rules remain to be investigated in the case when both vortices are fractional, i.e., $T_1=v_1{\bm S}_1\cdot{\bm \sigma}$ and $T_2=v_2{\bm S}_2\cdot{\bm \sigma}$. Inserting this into Eq.~(\ref{eq:two_paths}) one finds
\begin{equation}
T=v_1 v_2[{\bm S}_2\cdot(R^{-1}{\bm \sigma}R)]({\bm S}_1\cdot{\bm\sigma}).
\end{equation}
Using the property of group homomorphism, $R^{-1}{\bm \sigma} R=r{\bm\sigma}$, where $r{\bm S}_1={\bm S}_2$ we find
\begin{equation}
T=v_1 v_2 ({\bm S}_1\cdot{\bm \sigma})^2=v_1v_2\mathds{1}_{2\times2}.
\end{equation}
Comparing with Eq.~(\ref{eq:t}), we see that in all cases the index $v$ is multiplicative,
\begin{equation}
v=v_1v_2.
\end{equation}

\section{Homotopy Computation}\label{sec:homotopy_computation}

\subsection{Order-parameter manifold}
\label{subsec:order_parameter_manifold}

We restrict attention to local low-energy spiral textures for which the
principal eigenvector of $G_{\mu\nu}$ is well defined and the momentum
magnitude remains close to the spiral-ring radius $q_\circ$. The wave-vector
sector is then characterized by the orientation
\begin{equation}
    \hat{\boldsymbol q}(\theta)
    =\frac{\boldsymbol q}{|\boldsymbol q|}
    =(\cos\theta,\sin\theta),
    \qquad
    \theta\in\mathbb{R}/2\pi\mathbb{Z},
    \label{eq:qhat_parameterization}
\end{equation}
which takes values in $S^1_{\hat{\boldsymbol q}}\simeq U(1)$. This
$U(1)$ is an effective low-energy degree of freedom associated with the
continuous spiral ring; it need not correspond to an exact microscopic
symmetry away from the low-energy manifold.

The spin sector is described by the oriented orthonormal frame
\begin{equation}
    F
    =\bigl(
        \boldsymbol S,
        \boldsymbol\Omega_{\boldsymbol q},
        \boldsymbol S\times\boldsymbol\Omega_{\boldsymbol q}
      \bigr)
    \in \mathrm{SO}(3),
    \label{eq:local_spin_frame}
\end{equation}
where the three vectors are understood as the columns of $F$.
A global spin rotation $R\in\mathrm{SO}(3)$ acts by left multiplication,
$F\mapsto RF$, so the frame transforms as a rigid body. Before accounting
for redundancies, a local spiral would therefore be represented by a pair
\begin{equation}
    \bigl(\hat{\boldsymbol q}(\theta),F\bigr)
    \in S^1_{\hat{\boldsymbol q}}\times\mathrm{SO}(3).
    \label{eq:product_representative}
\end{equation}

This product representation is, however, two-to-one. The principal
eigenvector of $G_{\mu\nu}$ is defined only up to sign. Moreover, because
the spiral normal is defined using a directional derivative along
$\hat{\boldsymbol q}$,
\begin{equation}
    \boldsymbol\Omega_{\boldsymbol q}
    \propto
    \boldsymbol S\times
    \bigl(\hat{\boldsymbol q}\cdot\boldsymbol\nabla\bigr)\boldsymbol S,
    \label{eq:normal_direction_sign}
\end{equation}
the replacement $\hat{\boldsymbol q}\mapsto-\hat{\boldsymbol q}$ must be
accompanied by
$\boldsymbol\Omega_{\boldsymbol q}\mapsto-\boldsymbol\Omega_{\boldsymbol q}$.
The spin field itself is unchanged. Introduce the body-fixed rotation
\begin{equation}
    \mathsf D
    =R_{\hat{\boldsymbol e}_1}(\pi)
    =\operatorname{diag}(1,-1,-1),
    \qquad
    \mathsf D^2=\mathbf 1_3,
    \label{eq:body_fixed_D}
\end{equation}
where $\hat{\boldsymbol e}_1$ is the first axis of the body frame. Right
multiplication gives
\begin{equation}
    F\mathsf D
    =\bigl(
        \boldsymbol S,
        -\boldsymbol\Omega_{\boldsymbol q},
        -\boldsymbol S\times\boldsymbol\Omega_{\boldsymbol q}
      \bigr),
    \label{eq:frame_D_action}
\end{equation}
which is a rotation by $\pi$ about the physical spin axis
$\boldsymbol S$. Thus the two representatives
\begin{equation}
    \bigl(\hat{\boldsymbol q}(\theta),F\bigr)
    \quad\text{and}\quad
    \bigl(\hat{\boldsymbol q}(\theta+\pi),F\mathsf D\bigr)
    \label{eq:equivalent_representatives}
\end{equation}
describe the same local spiral configuration.

Defining the order-two action
\begin{equation}
    \gamma:(\theta,F)
    \longmapsto
    (\theta+\pi,F\mathsf D),
    \qquad
    \gamma^2=\mathrm{id},
    \label{eq:gamma_action}
\end{equation}
the physical order-parameter manifold is
\begin{equation}
    \mathcal M
    =
    \frac{S^1_{\hat{\boldsymbol q}}\times\mathrm{SO}(3)}
         {\langle\gamma\rangle}
    \simeq
    \frac{S^1_{\hat{\boldsymbol q}}\times\mathrm{SO}(3)}{\mathbb Z_2},
    \label{eq:order_parameter_manifold}
\end{equation}
where the explicit action in Eq.~\eqref{eq:gamma_action} is part of the
definition of the quotient. The action is free, so $\mathcal M$ is a
smooth manifold. Crucially, a path for which the momentum direction
changes by only $\pi$ can nevertheless close in $\mathcal M$, provided
that the frame simultaneously changes from $F$ to $F\mathsf D$. This is
the geometric origin of the fractional momentum vortices discussed
below.

\subsection{Homotopy classification and vortex charges}
\label{subsec:homotopy_classification}

In two spatial dimensions, a spin texture on a counterclockwise loop
$\mathcal C$ surrounding an isolated defect defines a closed path in
$\mathcal M$. In general, point defects are classified by conjugacy
classes of $\pi_1(\mathcal M)$; the fundamental group found below is
Abelian, so its conjugacy classes are its individual elements.

The fundamental group is most transparently obtained from the universal
cover
\begin{equation}
    \widetilde{\mathcal M}
    =\mathbb R\times\mathrm{SU}(2).
    \label{eq:universal_cover_space}
\end{equation}
Let $\rho:\mathrm{SU}(2)\rightarrow\mathrm{SO}(3)$ be the standard
double-covering map and define
\begin{equation}
    p:\widetilde{\mathcal M}\longrightarrow\mathcal M,
    \qquad
    p(\vartheta,U)
    =
    \Bigl[
        \bigl(\hat{\boldsymbol q}(\vartheta),\rho(U)\bigr)
    \Bigr]_{\mathcal M}.
    \label{eq:universal_cover_map}
\end{equation}
Here $\vartheta\in\mathbb R$ is the lifted momentum angle. A convenient
lift of $\mathsf D$ is
\begin{equation}
    \widetilde{\mathsf D}
    =\exp\!\left(-\frac{i\pi}{2}\sigma_1\right)
    =-i\sigma_1,
    \qquad
    \rho(\widetilde{\mathsf D})=\mathsf D,
    \qquad
    \widetilde{\mathsf D}^{\,2}=-\mathbf 1_2,
    \label{eq:D_lift}
\end{equation}
where $\sigma_a$ are the Pauli matrices. The full deck-transformation
group of $p$ is generated by
\begin{align}
    a:(\vartheta,U)
    &\longmapsto
    (\vartheta+\pi,U\widetilde{\mathsf D}),
    \label{eq:deck_a}
    \\
    z:(\vartheta,U)
    &\longmapsto
    (\vartheta,-U).
    \label{eq:deck_z}
\end{align}
The generator $a$ lifts the diagonal identification
\eqref{eq:gamma_action}, whereas $z$ is the nontrivial deck
transformation of the double cover
$\mathrm{SU}(2)\rightarrow\mathrm{SO}(3)$. They satisfy
\begin{equation}
    az=za,
    \qquad
    z^2=\mathrm{id},
    \label{eq:deck_relations_basic}
\end{equation}
while $a$ has infinite order because it shifts $\vartheta$ by $\pi$.
For reference,
\begin{align}
    a^2&:(\vartheta,U)
    \longmapsto
    (\vartheta+2\pi,-U),\\
    \qquad
    t\equiv a^2z&:
    (\vartheta,U)\longmapsto(\vartheta+2\pi,U),
    \label{eq:pure_momentum_deck_transformation}
\end{align}
so $t$ is the deck transformation associated with a pure $2\pi$
momentum winding. Since $\widetilde{\mathcal M}$ is simply connected,
its deck-transformation group is isomorphic to the fundamental group of
$\mathcal M$:
\begin{equation}
    \pi_1(\mathcal M)
    \cong
    \left\langle
        a,z\,\middle|\,[a,z]=1,\ z^2=1
    \right\rangle
    \cong\mathbb Z\times\mathbb Z_2.
    \label{eq:fundamental_group}
\end{equation}

We now translate this abstract group into labels adapted to the physical
vortex textures. Choose a lift $(\vartheta_0,U_0)$ of the order parameter
at the base point of $\mathcal C$. The lift of the path around a defect
ends at a deck-related point,
\begin{equation}
    (\vartheta_1,U_1)
    =a^u z^\eta(\vartheta_0,U_0)
    =\left(
        \vartheta_0+u\pi,
        U_0\widetilde{\mathsf D}^{\,u}(-1)^\eta
      \right),
    \label{eq:lifted_path_endpoint}
\end{equation}
with
\begin{equation}
    u\in\mathbb Z,
    \qquad
    \eta\in\{0,1\}.
    \label{eq:u_eta_values}
\end{equation}
The integer $u$ measures the winding of the momentum direction in units
of $\pi$,
\begin{equation}
    \Delta_{\mathcal C}\vartheta=u\pi.
    \label{eq:u_definition}
\end{equation}
Accordingly, $u=\pm2$ describes a conventional $\pm2\pi$ momentum
vortex or antivortex, whereas $u=\pm1$ describes a half-momentum vortex
or antivortex. The exponent $\eta$ is the independent $\mathbb Z_2$
label generated by $z$.

It is useful to encode the same information in the accumulated
$\mathrm{SU}(2)$ spin rotation
\begin{equation}
    T_{\mathcal C}=U_1U_0^{-1}
    =(-1)^\eta
      U_0\widetilde{\mathsf D}^{\,u}U_0^{-1}.
    \label{eq:T_C_definition}
\end{equation}
Let $\boldsymbol S_0$ be the spin at the base point. Because the first
column of $F_0=\rho(U_0)$ is $\boldsymbol S_0$,
\begin{equation}
    U_0\sigma_1U_0^{-1}
    =\boldsymbol S_0\cdot\boldsymbol\sigma,
    \qquad
    \boldsymbol\sigma=(\sigma_1,\sigma_2,\sigma_3).
    \label{eq:basepoint_spin_SU2}
\end{equation}
Using Eq.~\eqref{eq:D_lift}, Eq.~\eqref{eq:T_C_definition} becomes
\begin{equation}
    T_{\mathcal C}
    =
    \begin{cases}
        v\,\mathbf 1_2,
        & u\ \text{even},\\[2mm]
        v\,\boldsymbol S_0\cdot\boldsymbol\sigma,
        & u\ \text{odd},
    \end{cases}
    \qquad
    v=(-1)^\eta(-i)^u.
    \label{eq:v_definition_main_text}
\end{equation}
The physically convenient labels $(u,v)$ therefore obey
\begin{equation}
    u\ \text{even}
    \quad\Longleftrightarrow\quad
    v=\pm1,
    \qquad
    u\ \text{odd}
    \quad\Longleftrightarrow\quad
    v=\pm i,
    \label{eq:uv_parity_constraint}
\end{equation}
or, equivalently,
\begin{equation}
    v^2=(-1)^u.
    \label{eq:uv_compact_constraint}
\end{equation}
The values $v=\pm i$ are labels of the $\mathrm{SU}(2)$ lift and are not
complex-valued local observables. Appendix~\ref{sec:topological_index} shows explicitly that $v$
is independent of the choice of base point and is invariant under smooth
deformations of $\mathcal C$ that do not cross another defect.

For even $u$, both $\hat{\boldsymbol q}$ and $F$ return to their initial
representatives in the product space
$S^1_{\hat{\boldsymbol q}}\times\mathrm{SO}(3)$. The value $v=+1$
corresponds to a contractible spin-frame loop, whereas $v=-1$ is the
nontrivial element of
$\pi_1(\mathrm{SO}(3))=\mathbb Z_2$. In particular,
\begin{equation}
    (u,v)=(0,-1)
\end{equation}
is a pure $\mathbb Z_2$ spin vortex: the frame traces a noncontractible
closed path in $\mathrm{SO}(3)$, which can be represented by a net
$2\pi$ rotation about any axis. Different choices of the rotation axis
are homotopic. The sectors $(u,v)=(\pm2,+1)$ are pure full momentum
vortices and antivortices, while $(\pm2,-1)$ additionally carry the
nontrivial spin $\mathbb Z_2$ charge.

For odd $u$, the representative pair changes according to
\begin{equation}
    (\theta,F)
    \longrightarrow
    (\theta+\pi,F\mathsf D).
    \label{eq:odd_u_endpoint_product_space}
\end{equation}
The initial and final representatives are distinct in the product space
but identical in the quotient $\mathcal M$, so the path is closed only
because of the diagonal identification. The elementary sectors
$u=\pm1$ are therefore fractional momentum vortices. For each odd $u$,
the two values $v=\pm i$ represent the two inequivalent lifts of the same
$\pi$ rotation in $\mathrm{SO}(3)$. At a fixed base point they may be
represented as $+\pi$ and $-\pi$ rotations about the spin axis
$\boldsymbol S_0$. These paths have the same endpoint in
$\mathrm{SO}(3)$ but differ by a noncontractible $2\pi$ loop. With the
convention \eqref{eq:D_lift}, the $+\pi$ and $-\pi$ rotations carry
$v=-i$ and $v=+i$, respectively. We refer to this binary distinction as
the chirality of a fractional vortex.

The complete charge set can thus be written as
\begin{equation}
    \pi_1(\mathcal M)
    \simeq
    \left\{
        (u,v)\,\middle|\,
        u\in\mathbb Z,
        \ v\in\{\pm1,\pm i\},
        \ v^2=(-1)^u
    \right\},
    \label{eq:charge_set_uv}
\end{equation}
with composition law
\begin{equation}
    (u_1,v_1)(u_2,v_2)
    =
    (u_1+u_2,v_1v_2).
    \label{eq:charge_group_law}
\end{equation}
This law follows either from concatenating lifted loops or directly from
multiplying the corresponding deck transformations, and it will be used
to derive the fusion rules. Although the
local frame takes values in the non-Abelian group $\mathrm{SO}(3)$, the
fully resolved topological charge group in
Eq.~\eqref{eq:fundamental_group} is Abelian.

\begin{figure}
    \centering
    \includegraphics[width=\linewidth]{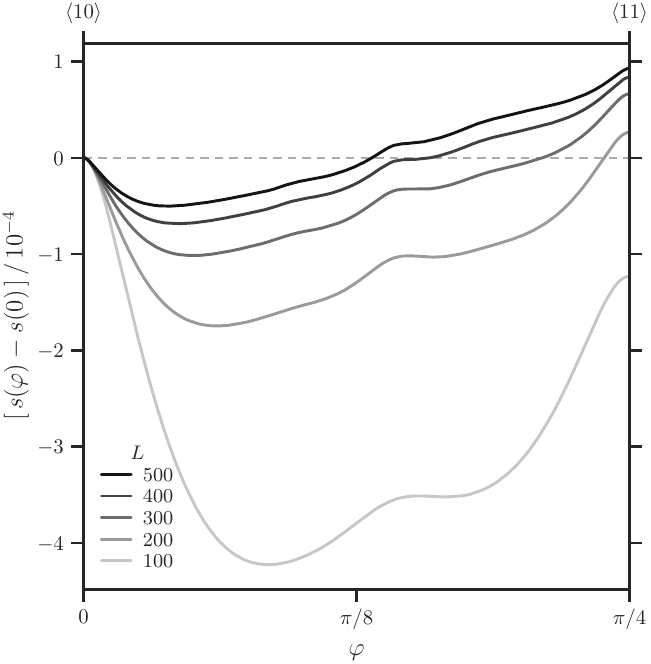}
    \caption{Spin-wave entropy per site $s$ for the different single-${\bm q}$ states along the ring-minimum. $\varphi$ parameterizes the direction of ${\bm q}$, with $\varphi=0$ corresponding to ${\bm q}$ lying in the $q_x$-direction. $L$ is the linear system size.}
    \label{fig:spinwaveentropy}
\end{figure}

\section{Linear-spin-wave entropy}\label{sec:lsw_entropy}

For low temperatures, the order is a single-domain single-${\bm q}$ spiral state. Here, entropy leads to the selection of a particular broken symmetry ground state at small, but finite, temperature by the order-by-disorder scenario~\cite{Villain1980, Henley1989, Chandra1990}. We have calculated the spin-wave entropy associated with the different ordering wave vectors along the ring minimum using linear spin-wave theory~\cite{Glittum2021, Seabra2016}. The result for $\delta = 0.03$ is shown in Fig.~\ref{fig:spinwaveentropy} for various linear system sizes $L$. The different single-${\bm q}$ states are parametrized by the direction $\varphi$ of ${\bm q}$, corresponding to the angle with the $q_x$-direction. For $L \geq 200$, the spin-wave entropy is highest for single-${\bm q}$ spiral states with ${\bm q}$ pointing along the lattice-diagonals $\langle 11 \rangle$.

\end{document}